\documentclass[showkeys,aps]{revtex4}

\usepackage{graphicx}
\usepackage{amsfonts}
\usepackage{amssymb}
\usepackage{amsmath}
\usepackage{hyperref}
\usepackage{natbib}
\usepackage{float}
\usepackage{dcolumn}
\usepackage{bm}
\usepackage{latexsym,color}

\begin{document}

\def\be{\begin{equation}}
\def\ee{\end{equation}}
\def\bea{\begin{eqnarray}}
\def\eea{\end{eqnarray}}

\def\mnras{Mon.~Not. Roy. Astr. Soc.}
\title{The Erez--Rosen Solution versus the~Hartle--Thorne~Solution}

\author{Kuantay Boshkayev $^{1,2}$, Hernando Quevedo $^{1,3,4}$, \\
Gulmira Nurbakyt $^{1}$, Algis Malybayev $^{1}$ and Ainur Urazalina $^{1}$}

\email{kuantay@mail.ru, quevedo@nucleares.unam.mx, \\
gumi-nur@mail.ru,  algis\_malybayev@mail.ru, y.a.a.707@mail.ru}

\affiliation{
$^{1}$ National Nanotechnology Laboratory of Open Type, Department of Theoretical and Nuclear Physics, Al-Farabi Kazakh National University,  Almaty 050040, Kazakhstan;  \\
$^{2}$ Energetic Cosmos Laboratory, Department of Physics, Nazarbayev University, Qabanbay~Batyr~53, Astana 010000, Kazakhstan.\\
$^{3}$ Dipartimento di Fisica and ICRA, Universit\`a di Roma ``La Sapienza'', I-00185 Roma, Italy\\
$^{4}$ Instituto de Ciencias Nucleares, Universidad Nacional Aut\'onoma de M\'exico, AP 70543, Mexico city DF 04510, Mexico
}

\begin{abstract}
In this work, we investigate the correspondence between the Erez--Rosen and Hartle--Thorne solutions. We explicitly show how to establish the relationship and find the coordinate transformations between the two metrics. For this purpose the two metrics must have the same approximation and describe the gravitational field of static objects. Since both the Erez--Rosen and the Hartle--Thorne solutions are particular solutions of a more general solution, the Zipoy--Voorhees transformation is applied to the exact Erez--Rosen metric in order to obtain a generalized solution in terms of the Zipoy--Voorhees parameter $\delta=1+sq$. The Geroch--Hansen multipole moments of the generalized Erez--Rosen metric are calculated to find the definition of the total mass and quadrupole moment in terms of the mass $m$, quadrupole $q$ and Zipoy--Voorhees $\delta$ parameters. The coordinate transformations between the metrics are found in the approximation of $\sim$q.  It is shown that the Zipoy--Voorhees parameter is equal to $\delta=1-q$ with $s=-1$. This result is in agreement with previous results in the literature.
\end{abstract} 


\maketitle

\section{Introduction}

There exists quite a large number of exact and approximate solutions of the Einstein field equations (EFE) in the literature \cite{stefanietal, bqr12}. We mainly focus here only on two exterior solutions: The exact Erez--Rosen (ER) \cite{erro59} and approximate Hartle--Thorne (HT) \cite{hartle1967, ht1968} solutions. Both of them describe the gravitational field of astrophysical objects. Though the ER metric is exact and describes only the exterior part of the static deformed object, in turn, the HT metric is approximate and  can be used to investigate both interior and exterior fields of slowly rotating and slightly deformed astrophysical objects in the strong field regime. In this regard, it is interesting to show the relationship between these solutions in the limiting static case with a small deformation.

Erez and Rosen obtained their solution in 1959 \cite{erro59} by using the Weyl method \cite{weyl}. This metric was also analyzed by applying the spheroidal coordinates, which are adapted to characterize the gravitational field of non-spherically symmetric bodies. The original solution contained some typos and misprints, which were later corrected in several numerical coefficients by Doroshkevich (1966)~\cite{doretal1966}, Winicour et al. (1968) \cite{winietal1968} and Young and Coulter (1969) \cite{yc1969}. The physical properties of the ER metric were investigated by Zeldovich and Novikov \cite{zn71} and later by Quevedo and Parkes \cite{qp1989}. More~general solutions involving multipole moments were obtained by Quevedo~\cite{quev89, quev90}, Quevedo and Mashhoon (QM)~\cite{qm1985, qm1990}.

The QM solution is an exact exterior metric describing the gravitational field of a rotating deformed mass \cite{qm1990}, which is a stationary axisymmetric solution of the vacuum Einstein equations belonging to the class of Weyl--Lewis--Papapetrou \cite{weyl, lewis, pap}. The QM solution involving only the mass parameter $m$, quadrupole parameter $q$ and rotation parameter $a$ (angular momentum per unit mass)  is a generalization of the Kerr metric \cite{kerr63}, so it reduces to the exact Kerr solution when the quadrupole parameter vanishes $q\rightarrow0$, and to the ER spacetime when the rotation parameter vanishes $a\rightarrow0$ \cite{bglq09}. It has been also shown that the general form of the QM solution with the Zipoy--Voorhees parameter in the limiting case is equivalent to the exterior HT solution up to the first order in the quadrupole parameter $q$ and to the second order terms in the rotation parameter $a$ \cite{bqr12, bglq09}.

Hartle developed his formalism in order to investigate the physical properties of slowly rotating relativistic stars in his pioneering paper in 1967 \cite{hartle1967}. All physical quantities describing the equilibrium configurations of rotating stars such as the change in mass, gravitational potential, eccentricity, binding energy, quadrupole moment, etc., were proportional to the square of the star's angular velocity  $\Omega^2$. Hartle and Thorne tested the formalism for different equations of state of relativistic objects~\cite{ht1968}. Since~then the solution has been known as the Hartle--Thorne solution in the literature. Unlike~well-known exact solutions, the HT solution possesses an internal counterpart \cite{hartle1967, ster03}, which makes it more practical in the astrophysical context to investigate the equilibrium structure and physical characteristics of relativistic compact objects such as white dwarfs, neutron stars and hypothetical quark stars \cite{brrs2013, bbrr2014, gw1992, ums2013}. Recently, the HT metric has been extended up to an $\Omega^4$ approximation \cite{ykpya2014}.

The purpose of this work is to find the relationship between the ER and HT solutions and show their equivalence in the limiting static case with a small deformation. The signature of the line elements throughout this article is adopted as (-- + + +) and the geometrical units are used ($G=c=1$).

It should be emphasized that the relationship between the ER and HT solutions was established by Mashhoon and Theiss in 1991 \cite{mth1991}. However, in this work we derive the same result in an instructive way, providing all technical details. The paper pursues pure scientific and academic purposes.

The work is organized as follows. We review the main properties of the ER solution in Section \ref{ER}. 
The linearized, up to the first order in $q$, Erez--Rosen solution in terms of the Zipoy--Voorhees parameter $\delta=1+sq$  is considered in Section \ref{lER}. The main physical characteristics of the exterior Hartle--Thorne solution are discussed in Section \ref{HT}. Using the perturbation method, the coordinate transformations are sought in Section \ref{CT}. Finally, we summarize our conclusions and discuss future prospects.


\section{The Erez--Rosen Metric}\label{ER}

The Erez--Rosen metric is an exact exterior solution with mass and quadrupole parameters that describes the gravitational field of static deformed objects in the strong field regime \cite{qta2012}. It belongs to the Weyl class of static axisymmetric vacuum solutions in prolate spheroidal coordinates $(t,x,y,\varphi)$, with $x\geq 1$  and $-1\leq y \leq1$

\be \label{lel}
 ds^2 =-e^{2\psi} dt ^2 + m^2 e^{-2\psi}\left[ e^{2\gamma}(x^2-y^2)\left( \frac{dx^2}{x^2-1} + \frac{dy^2}{1-y^2} \right) + (x^2-1)(1-y^2) d\varphi^2\right] \ ,
\ee
where the metric functions $\psi$ and $\gamma$ depend on the spatial coordinates $x$ and $y$, only, and $m$ represents the mass parameter.

The solution found by Erez and Rosen has the following form \cite{quev12}
\bea
\psi  & = & \frac{1}{2}\ln\left(\frac{x-1}{x+1}\right) + \frac{1}{2}q(3y^2-1) \left[\frac{1}{4}(3x^2-1)\ln\left(\frac{x-1}{x+1}\right)+\frac{3}{2} x\right] \,
\eea
and
\bea
\gamma = & &\frac{1}{2}(1+q)^2 \ln\left( \frac{x^2-1}{x^2-y^2}\right) - \frac{3}{2} q(1-y^2)\left[x \ln\left(\frac{x-1}{x+1}\right) + 2\right]\nonumber\\
& + & \frac{9}{16}q^2(1-y^2)\bigg[ x^2+4y^2 - 9 x^2y^2-\frac{4}{3} 
+ x\left(x^2+7y^2-9x^2y^2-\frac{5}{3}\right)\ln\left(\frac{x-1}{x+1}\right) \\
& + & \frac{1}{4}(x^2-1)(x^2+y^2-9x^2y^2-1)\ln^2\left(\frac{x-1}{x+1}\right)\bigg]\ . \nonumber
\eea 

However, nowadays, it is considered to be one particular type of a more general class of solutions of the field equations.


To generate a more general solution the Zipoy--Voorhees \cite{zip66, voor70} transformation is applied to the Erez--Rosen metric with line element (\ref{lel}) 
as $\psi\rightarrow \delta\psi$ and $\gamma\rightarrow \delta^2 \gamma$. 
For this new generalized Erez--Rosen solution with the Zipoy--Voorhees parameter $\delta$, the Geroch--Hansen multiples \cite{ger, hans, quev89, quev90, ernst, fqs2018, fs2018} are given~by \bea
&&M_{2k+1}=0, \qquad k=0, 1, 2, ... ,\\
&&M_0=m\delta, \qquad M_2=\frac{1}{15}m^3\delta(5+2q-5\delta^2), ...
\eea

Furthermore, if we adopt that $\delta=1+sq$, then
\be\label{eqmm15}
M_0=m(1+sq), \qquad M_2=\frac{1}{15}m^3q(1+sq)(2-10s-5s^2q), ...
\ee

For vanishing $s=0$ or, equivalently, $\delta=1$ we obtain the multipole moments of the original Erez--Rosen solution.
If we retain only linear terms in the quadrupole parameter $q$ then (\ref{eqmm15}) becomes
\be\label{eqmm16}
M_0=m(1+sq), \qquad M_2=\frac{2}{15}m^3q(1-5s), ...
\ee


\section{The Linearized Erez--Rosen Solution in Terms of the Zipoy--Voorhees Parameter}\label{lER}

The new generalized Erez--Rosen metric with $\delta$ is linearized up to the first order in $q$ and the final result is written in spherical-like coordinates $x=r/m-1$ and $y=\cos\theta$
\begin{equation}\label{lelzv}
\begin{split}
ds^2&= -\left(1-\frac{2m}{r}\right)\left[1+2q(\psi_1+s\psi_0)\right]dt^2 \\ 
&+\left[1+2q(\gamma_1-\psi_1+2s\gamma_0-s\psi_0)\right]\left(\frac{dr^2}{1-\frac{2m}{r}}+r^2d\theta^2\right)+\left[1-2q(\psi_1+s\psi_0)\right]r^2\sin^2\theta d\phi^2 \ , 
\end{split}
\end{equation}
where
\bea
\psi_0 &=& \frac{1}{2}\ln\left(\frac{x-1}{x+1}\right),  \quad \psi_1 = \frac{1}{2}(3y^2-1) \left[\frac{1}{4}(3x^2-1)\ln\left(\frac{x-1}{x+1}\right)+\frac{3}{2} x\right] \ ,
\eea
and
\bea
\gamma_0 &=& \frac{1}{2} \ln\left( \frac{x^2-1}{x^2-y^2}\right), \quad  \gamma_1 = \ln\left( \frac{x^2-1}{x^2-y^2}\right)- \frac{3}{2}(1-y^2)\left[x \ln\left(\frac{x-1}{x+1}\right) + 2\right] .
\eea

In the limiting case $q=0$, we recover from here 
the Schwarzschild solution:
\be
ds^2=-\left(1-\frac{2m}{r}\right)dt^2+\frac{dr^2}{1-\frac{2m}{r}}+r^2d\theta^2+r^2\sin^2\theta d\phi^2 \ .
\ee

Consequently, the linearized Equation (\ref{lelzv}) can be considered as a generalization of the Schwarzschild metric, which includes the quadrupole contribution of the Erez--Rosen solution up to the first order in $q$.


\section{The Exterior Hartle--Thorne Solution}\label{HT}
The general form of the exterior approximate HT metric \cite{hartle1967, ht1968} in spherical ($t, R, \Theta, \phi$) coordinates is given by
\begin{align}\label{ht1}
ds^2&=-\left(1-\frac{2{ M }}{R}\right)\bigg[1+2k_1P_2(\cos\Theta)
-2\left(1-\frac{2{ M}}{R}\right)^{-1}\frac{J^{2}}{R^{4}}(2\cos^2\Theta-1)\bigg]dt^2\nonumber\\
&+\left(1-\frac{2{ M}}{R}\right)^{-1}\bigg[1-2\left(k_1-\frac{6 J^{2}}{R^4}\right)P_2(\cos\Theta)-2\left(1-\frac{2{ M}}{R}\right)^{-1}\frac{J^{2}}{R^4}\bigg]dR^2 \\
&+R^2[1-2k_2P_2(\cos\Theta)](d\Theta^2+\sin^2\Theta d\phi^2) -\frac{4J}{R}\sin^2\Theta dt d\phi \,\nonumber
\end{align}
where
\begin{align*}
k_1&=\frac{J^2}{M R^3}\left(1+\frac{M}{R}\right)+\frac{5}{8}\frac{Q-J^{2}/{M}}{M^3}Q_2^2(x) \;,\\
k_2&=k_1+\frac{J^{2}}{R^4}+\frac{5}{4}\frac{Q-J^{2}/{ M}}{M^2R}\left(1-\frac{2M}{R}\right)^{-1/2}Q_2^1(x) \;,
\end{align*}
and
\begin{align*}
\label{legfunc}
Q_2^1(x)&=(x^2-1)^{1/2}\left[\frac{3x}{2}\ln\left(\frac{x+1}{x-1}\right)-\frac{3x^2-2}{x^2-1}\right] \;, \\
Q_2^2(x)&=(x^2-1)\left[\frac{3}{2}\ln\left(\frac{x+1}{x-1}\right)-\frac{3x^3-5x}{(x^2-1)^2}\right] \;,
\end{align*}
\noindent are the associated Legendre functions of the second kind, being $P_2(\cos\Theta)=(1/2)(3\cos^2\Theta-1)$ the Legendre polynomial, and $x=R/M -1$. This form of the metric is known in the literature as the Hartle--Thorne metric. The constants $M$, $J$  and $Q$  are the total mass, angular momentum and mass quadrupole moment of the rotating object, respectively. To obtain the exact numerical values of $M$, $J$ and $Q$, the exterior and interior line elements have to be matched at the surface of the star \cite{ht1968, bglq09, bbrs2013}.

For vanishing angular momentum the HT solution reduces to
\begin{equation}
\begin{split}
ds^2&=-\left(1-\frac{2{ M }}{R}\right)\bigg[1+2k_1P_2(\cos\Theta)\bigg]dt^2+\left(1-\frac{2{ M}}{R}\right)^{-1}\bigg[1-2k_1P_2(\cos\Theta)\bigg]dR^2\\
&+R^2[1-2k_2P_2(\cos\Theta)](d\Theta^2+\sin^2\Theta d\phi^2) \ ,
\end{split}
\end{equation}
where $k_1$ and $k_2$ now are
\be
k_1=\frac{5}{8}\frac{Q}{M^3}Q_2^2(x) \ , \qquad k_2=k_1+\frac{5}{4}\frac{Q}{ M^2R}\left(1-\frac{2M}{R}\right)^{-1/2}Q_2^1(x) \;,
\ee

To find the relationship with the ER solution, we will use Equation (\ref{lelzv}).
The Geroch--Hansen multipole moments for the HT metric are
\be\label{mmmq}
M_0=M, \qquad M_2=-Q, ...
\ee
where $Q$ has positive sign for oblate objects and negative for prolate objects, according to Hartle's~definition.


\section{Coordinate Transformations}\label{CT}

To obtain the correspondence between the  ER solution, with coordinates ($t, r, \theta, \phi$), and the HT solution, with coordinates ($t, R, \Theta, \phi$), both solutions must be written in the same coordinates. We~search for a coordinate transformation of the following form

\be
r \rightarrow R+q f_1(R,\Theta) \, \qquad \theta \rightarrow \Theta+q f_2(R,\Theta)
\ee
where $f_1(R,\Theta)$ and $f_2(R,\Theta)$ are the unknown  functions.
The total differentials of the coordinates are given by
\bea\label{eqct}
dr &=& \frac{\partial r}{\partial R}dR+\frac{\partial r}{\partial \Theta}d\Theta=\left(1+q \frac{\partial f_1(R,\Theta) }{\partial R}\right)dR +q\left(\frac{\partial f_1(R,\Theta)}{\partial \Theta}\right)d\Theta \ , \\
d\theta &=& \frac{\partial \theta}{\partial R}dR+\frac{\partial \theta}{\partial \Theta}d\Theta=q\left(\frac{\partial f_2(R,\Theta)}{\partial R}\right)dR+\left(1+q \frac{\partial f_2(R,\Theta) }{\partial \Theta}\right)d\Theta\ .
\eea

These expressions should be plugged into  Equation (\ref{lelzv}), taking into account Equations~(\ref{eqmm16})~and~(\ref{mmmq}),~i.e.,
\be
m=M(1-sq) \ , \qquad q=-\frac{15}{2(1-5s)}\frac{Q}{M^3} \, . 
\ee

Then, only linear terms in the quadrupole moment $Q$  must be retained. We thus obtain the linearized ER metric in the same coordinates as the HT solution. Furthermore, if we compare the components of the metric tensor $g_{tt}$   of the ER and HT solutions, we find the value of the sought function $f_1(R,\Theta)$  as
\be
f_1(R,\Theta)=f_{10}(R,\Theta)+(1+s)\left[f_{11}(R)+f_{12}(R,\Theta)\right] \ ,
\ee
where
\bea
f_{10}(R,\Theta) &=& M+\frac{3}{2}M\sin^2\Theta\left[\frac{R}{M}-1+\frac{R}{M}\left(\frac{R}{2M}-1\right)\ln\left(1-\frac{2M}{R}\right) \right] \ , \nonumber \\
f_{11}(R) &=& -M \left[\frac{5}{6}+\frac{8R}{3M}-\frac{15R^2}{4M^2}+\frac{5R^3}{4M^3}-\frac{R}{M}\left(1-\frac{3R}{M}+\frac{5R^2}{2M^2}-\frac{5R^3}{8M^3}\right)\ln\left(1-\frac{2M}{R}\right) \right] \ ,  \nonumber \\
f_{12}(R,\Theta) &=& M \sin^2\Theta\left[\frac{5}{4}+\frac{5R}{2M}-\frac{45R^2}{8M^2}+\frac{15R^3}{8M^3}+\frac{15R^2}{4M^2}\left(1-\frac{R}{M}+\frac{R^2}{4M^2}\right) \ln\left(1-\frac{2M}{R}\right) \right] \ . \nonumber
\eea

Analogously, by comparing only the azimuthal components of the metric tensor $g_{\phi\phi}$  of the ER and HT solutions, we find the value of the sought function $f_2(R,\Theta)$ as
\be
f_2(R,\Theta)=f_{20}(R,\Theta)+(1+s)\left[f_{21}(R)+f_{22}(R,\Theta)\right] \ ,
\ee
where
\bea
f_{20}(R,\Theta) &=& -\frac{3}{2}\cos\Theta \sin\Theta\left[2+\left(\frac{R}{M}-1\right)\ln\left(1-\frac{2M}{R}\right) \right] \ , \nonumber \\
f_{21}(R) &=& \frac{47}{12}-\frac{5R}{2M}+\frac{5R^2}{4M^2}+\left(-\frac{7}{4}+\frac{3R}{M}-\frac{15R^2}{8M^2}+\frac{5R^3}{8M^3}\right)\ln\left(1-\frac{2M}{R}\right)  \ , \nonumber\\
f_{22}(R,\Theta) &=& - \sin^2\Theta\left[\frac{35}{8}-\frac{15R}{4M}+\frac{15R^2}{8M^2}-\left(\frac{15}{8}-\frac{15R}{4M}+\frac{45R^2}{16M^2}-\frac{15R^3}{16M^3}\right) \ln\left(1-\frac{2M}{R}\right) \right] \ . \nonumber
\eea

We need to find also the exact value of $s$. To this end, we equate the mixed component of the metric tensor $g_{R\Theta}$  of the ER solution, written in  ($R,\Theta$) coordinates, to zero as it is absent in the HT solution. This gives the following condition
\be
\left(1-\frac{2M}{R}\right)^{-1}\frac{\partial f_1(R,\Theta)}{\partial \Theta}+R^2\frac{\partial f_2(R,\Theta)}{\partial R} = 0 \ .
\ee

As a result, we find from this condition that $s=-1$ . Correspondingly, the multipole moments will be
\be\label{eq:GH}
M=m(1-q) ,\, \qquad Q=-\frac{4}{5}m^3q \ ,  ...
\ee
and the coordinate transformations will have the following form
\bea
r &\rightarrow & R+q \left\{M+\frac{3}{2}M\sin^2\Theta\left[\frac{R}{M}-1+\frac{R}{M}\left(\frac{R}{2M}-1\right)\ln\left(1-\frac{2M}{R}\right) \right]\right\} \ , \label{coorR} \\
\theta &\rightarrow & \Theta-\frac{3}{2}q\cos\Theta \sin\Theta\left[2+\left(\frac{R}{M}-1\right)\ln\left(1-\frac{2M}{R}\right) \right]  \ . \label{coorTh}
\eea

These coordinate transformations were originally obtained by Mashhoon and Theiss \cite{mth1991}. Here, we simply reproduced their results including all intermediate calculations. 
In the limit of linear quadrupole moment, to our knowledge, the only way to find the relationship between the two solutions is to consider the Zipoy--Voorhees transformation. 
However, recently  Frutos-Alfaro and Soffel \cite{fs2017} showed that in the limit of $\sim M^2$ and $\sim Q$ (negleting also terms $\sim M^2Q$) the Zipoy--Voorhees transformation is not needed. They obtained the following transformations
\bea
r &\rightarrow & R-\frac{1}{9}\frac{M}{R^3}Q\left[5P_2^2-4P_2-1\right] \ , \label{rFS} \\
\theta &\rightarrow & \Theta+\frac{1}{6}\frac{M}{R^4}Q\left[5P_2-2\right]\cos\Theta \sin\Theta  \ , \label{thFS}
\eea
where 
$P_2=P_2(\cos\Theta)$ and
\be\label{eq:GHFS}
M=m ,\, \qquad Q=-\frac{2}{15}m^3q \ ,  ...
\ee

Obviously, Equations~\eqref{rFS} and \eqref{thFS} are different from Equations~\eqref{coorR} and \eqref{coorTh} even in the limiting case. { Indeed, there is no any agreement between Equations~\eqref{coorR} and~\eqref{coorTh} and Equations~\eqref{rFS} and~\eqref{thFS} in any given limit}. However, Equations~\eqref{rFS} and \eqref{thFS} are valid only in the limit of $\sim M^2$ and $\sim Q$ and Equations~\eqref{coorR} and \eqref{coorTh} are correct in the limit of $\sim Q$ when no other approximation is made. This means that the explicit form of the coordinate transformation depends upon the approximation, which is adopted in each particular case, and the use of the Zipoy--Voorhees transformation. { This can  be already seen from the definitions of the Geroch--Hansen multipole moments (see Equations~\eqref{eq:GH}~and~\eqref{eq:GHFS}).}


\section{Conclusions}

In this work, we investigated the  Erez--Rosen and  Hartle--Thorne metrics for small quadrupole moment and found their relationship in the absence of rotation by using the perturbation method. 
We~showed that the approximate Erez--Rosen line element coincides with the Hartle--Thorne solution, in the limit of $\sim q$, when the former is considered with a  Zipoy--Voorhees transformation. Accordingly, using the invariant definition proposed by Geroch and Hansen, we have shown that   the corresponding mass monopole and quadrupole moment coincide in this approximation as well. 

To find the explicit form of the coordinate transformation, we compared the metric functions and obtained that the Zipoy--Voorhees transformation should be subject to the condition $s=-1$. { We~showed that the condition $s=-1$ is a direct consequence of $g_{R\Theta}=0$, though the procedure to obtain this value was obscure and unclear in the literature}. Moreover, we found that the explicit form of the coordinate transformation depends entirely on the approximation which is used in each particular case. 

In view of the results obtained recently in \cite{bqaks2016, bggqt2016, allahyari2019, allahyari2019b}, it would be interesting to establish the relationship between the Erez--Rosen and q-metrics. This will be the issue of future studies.\vspace{6pt}

\section*{Acknowledgments} The authors thank the journal editorial board for the invitation to publish the paper and anonymous referees for useful comments and remarks, which improved the presentation of the work. This work was supported by the Ministry of Education and Science of the Republic of Kazakhstan, Program IRN: BR05236494 and Grant IRN: AP05135753.


\end{document}